\documentclass[10pt,aps,prl,showpacs,footinbib,twocolumn]{revtex4-1}
\usepackage[usenames,dvipsnames]{color}
\usepackage{amsfonts}
\usepackage{amssymb}
\usepackage{amsmath}
\usepackage{graphicx}
\usepackage{epstopdf}
\usepackage[colorlinks]{hyperref}
\usepackage{caption}
\usepackage{subcaption}
\usepackage[normalem]{ulem}
\newcommand{\red}[1]{{\color{red}#1}}

\begin{document}

\title{New class of quasinormal modes of neutron stars in scalar-tensor gravity}

\author{Raissa F.\ P.\ Mendes}
\email{rfpmendes@id.uff.br}
\affiliation{Instituto de F\'isica, Universidade Federal Fluminense, \\
Av.~Gal.~Milton Tavares de Souza s/n, Gragoat\'a, 24210-346 Niter\'oi, 
Rio de Janeiro, Brazil.}

\author{N\'estor Ortiz}
\email{nortiz@perimeterinstitute.ca}
\affiliation{Perimeter Institute for Theoretical Physics, 31 Caroline Street North, Waterloo, Ontario, N2L 2Y5, Canada.}

\date{\today}

\begin{abstract}
Detection of the characteristic spectrum of pulsating neutron stars can be a powerful tool not only to probe the nuclear equation of state but also to test modifications to general relativity. However, the shift in the oscillation spectrum induced by modified theories of gravity is often small and degenerate with our ignorance of the equation of state. In this Letter, we show that the coupling to additional degrees of freedom present in modified theories of gravity can give rise to new families of modes, with no counterpart in general relativity, which could be sufficiently well resolved in frequency space to allow for clear detection. We present a realization of this idea by performing a thorough study of radial oscillations of neutron stars in massless scalar-tensor theories of gravity. We anticipate astrophysical scenarios where the presence of this class of quasinormal modes could be probed with electromagnetic and gravitational wave measurements.
\end{abstract}

\pacs{04.50.Kd, 04.80.Cc} 

\maketitle

\noindent \textit{Introduction.---}The study of stellar pulsations has long provided a valuable tool to scrutinize the interior of stars~\cite{Unno1989}. With the detection of gravitational waves (GWs) by the LIGO and Virgo Collaborations~\cite{Abbott2016a,LIGOScientificCollaboration2016a,Abbott2017, LIGOScientificCollaboration2017,Abbott2017a}, the study of \textit{relativistic} pulsations of compact objects has finally met observations, giving rise to a wealth of possibilities to probe the workings of gravity in the strong field regime, from the geometry of black holes to the internal structure of relativistic stars~\cite{Nollert1999,Kokkotas1999,Berti2018,Berti2018a}.
In order to build intuition into how modifications to general relativity (GR) can alter the emission of GWs, it is insightful to consider concrete frameworks in which Einstein's theory can be generalized. A particular framework, subject of the present study, is that of scalar-tensor theories of gravity (STTs)~\cite{Damour1992,Fujii2003}. These theories are extensions to GR which include, as an additional mediator of the gravitational interaction, a (possibly self-interacting) scalar field $\phi$ that couples nonminimally to the metric sector. Interestingly, for some choices of the nonminimal coupling function, STTs predict a radically different phenomenology for relativistic stars, due to the existence of a nonperturbative strong field effect known as {\it spontaneous scalarization}~\cite{Damour1993,Damour1996,Salgado1998,Harada1998}.

Relativistic stars that undergo the process of spontaneous scalarization have their internal structure altered with respect to GR. Consequently, their vibration modes are also modified. Stellar pulsations in STTs have been studied in a variety of contexts, often using strong approximation schemes~\cite{Sotani2004,Sotani2005,Silva2014,Sotani2014,Yazadjiev2017}. Here we examine the problem of \textit{radial} oscillations of neutron stars (NSs) in STTs in its full generality.

In GR, the main interest in radial oscillations of relativistic stars lies in the information they provide about (in)stability against gravitational collapse \cite{Chandrasekhar1964,Kokkotas2001}. However, in STTs the scalar sector of the gravitational field is dynamical even in spherical symmetry, and can carry energy---and information---away from the source, which adds to the relevance of studying spherical perturbations.

Our analysis of the full radial problem reveals the existence of entirely new families of quasinormal modes (QNM), with no counterpart in GR, which could not be captured by the Cowling approximation used previously \cite{Sotani2014}.
Since these modes can be explicitly related to the scalar degree of freedom, we dub them {\it scalar} or $\phi$ modes.
Indeed, one could expect the coupling to new degrees of freedom to introduce new families of modes, in addition to just shifting the original spectrum. This is the case of stellar oscillations in GR, where the coupling to metric degrees of freedom  not only modifies the Newtonian modes but also introduces new {\it spacetime} or $w$ modes \cite{Kokkotas1986,Kokkotas1992a}. Interestingly, the new family of scalar modes reported in this Letter may be relatively long lived, and lie in a lower frequency range (of $\lesssim 1$ kHz) more accessible to current GW detectors. Also, the $\phi$-mode frequencies can be clearly distinguished from those of GR even if the coupling to the scalar field is small, {\it i.e.} for stars with small scalar charges, in which case one might expect the GR fluid modes to only acquire a fine structure (easily mimicked by a change in the nuclear equation of state). 
Our findings suggest that the existence of new families of stellar QNMs could constitute a clear fingerprint of theories of modified gravity containing additional (scalar, vector...) degrees of freedom and that looking for their signatures in astrophysical data could be a promising way to improve constraints on such theories.
\\
\newpage
\noindent\textit{Framework.---}We focus on a class of massless STTs described by the action (we use $c=G=1$ unless specified)
\begin{equation}\label{eq:action}
S = \! \frac{1}{16 \pi} \! \int \! d^4 x \sqrt{-g} \left( R - 2 \nabla_\mu \phi \nabla^\mu \phi \right) + S_m[\Xi_m; a(\phi)^2 g_{\mu\nu}],
\end{equation}
where $\Xi_m$ represents matter fields with action $S_m$ that couple to the conformally rescaled (Jordan-frame) metric $\tilde{g}_{\mu\nu} := a(\phi)^2 g_{\mu\nu}$. 
In the following, we consider two representative coupling models, 
\begin{align}
\textrm{{\bf Model 1} (M1): } &a(\phi) = \left[ \cosh \left(\sqrt{3} \beta \phi \right) \right]^{1/(3\beta)}, \label{eq:MO} \\
\textrm{{\bf Model 2} (M2): } &a(\phi) = e^{\beta \phi^2/2}, \label{eq:DEF}
\end{align}
where $\beta \in \mathbb{R}$. M1 is an analytical approximation to the coupling function of a ``standard'' massless scalar field nonminimally coupled to gravity (see Sec.~III~B in Ref.~\cite{Mendes2016}), whereas M2 is the simplest model leading to spontaneous scalarization~\cite{Damour1993}. Both models agree up to the cubic term in an expansion around the ``cosmological'' value $\phi = 0$ \footnote{The cosmological value of the scalar field is constrained by Solar System observations to be extremely small \cite{Bertotti2003}. For simplicity, we assume it to be zero.} and in this particular case have the same post-Newtonian expansion as GR~\cite{Damour1996a}.

The variation of Eq.~(\ref{eq:action}) yields the field equations
\begin{align}
  & G_{\mu\nu} - 2 \nabla_\mu \phi \nabla_\nu \phi + 
	g_{\mu\nu} \nabla_\sigma \phi 
	\nabla^\sigma \phi  = 8\pi a(\phi)^2 \tilde{T}_{\mu\nu} ,\label{eq:metric_eq}
	\\
  &	\nabla^\mu \nabla_\mu \phi = - 4 \pi a(\phi)^4 \alpha(\phi) \tilde{T},
	\label{eq:phi_eq}
\end{align} 
where $\alpha(\phi):= d\ln a(\phi)/d\phi$, 
$\tilde{T} := \tilde{g}_{\mu\nu} \tilde{T}^{\mu\nu}$, and $\tilde{T}^{\mu\nu} := (2/\sqrt{-\tilde{g}}) \delta S_m[\Xi, \tilde{g}_{\alpha\beta}]/\delta \tilde{g}_{\mu\nu}$ is the (Jordan-frame) stress-energy tensor of matter fields, which is covariantly conserved ($\tilde{\nabla}_\mu \tilde{T}^{\mu\nu} =0$). We model NSs by a perfect fluid stress-energy tensor $\tilde{T}^{\mu\nu} = (\tilde{\epsilon} + \tilde{p}) \tilde{u}^\mu \tilde{u}^\nu + \tilde{p}\tilde{g}^{\mu\nu}$, with a barotropic equation of state (EoS) relating the pressure $\tilde{p}$ and the rest-mass density $\tilde{\rho}$. The total energy density $\tilde{\epsilon}$ is obtained from the first law of thermodynamics, $d \tilde{\epsilon} = (\tilde{\epsilon} + \tilde{p}) d\tilde{\rho}/\tilde{\rho}$. Conservation of baryon number further implies $\tilde{\nabla}_\mu (\tilde{\rho} \tilde{u}^\mu) = 0$, where $\tilde{u}^\mu$ denotes the four-velocity of fluid elements.
For our purposes, it is enough to adopt a polytropic EoS, $\tilde{p}(\tilde{\rho}) = K \rho_0 (\tilde{\rho}/\rho_0)^\gamma$, consisting of two phases, a stiff core ($\gamma_1=3$) and a soft crust ($\gamma_2 =1.3$). We choose $K_1=0.005$ and the transition density to be $\tilde{\rho}_t = 0.1 \rho_0$, where $\rho_0 = 1.66 \times 10^{14}$g/cm$^3$ \footnote{In spite of its pragmatical simplicity, this EoS is consistent with astrophysical constraints~\cite{Ozel2016}. Moreover, $\gamma =3$ is a typical effective polytropic exponent for the core of NSs~\cite{Read2009}, and our chosen value for $K_1$ guarantees a maximum mass consistent with observations~\cite{Antoniadis2013}. Further discussion on these choices can be found in Sec.~IIIA of Ref.~\cite{Mendes2016}.}.
\\

\noindent \textit{Background.---}The unperturbed equilibrium star is assumed to be static and spherically symmetric, with a line element of the form
\begin{equation} \label{eq:metric}
ds^2 = - e^{2 \nu} dt^2 + e^{2\lambda} dr^2 + r^2 
\left( d\vartheta^2 + \sin^2\vartheta d\varphi^2 \right),
\end{equation}
where $\nu = \nu(r)$, $\lambda = \lambda(r)$. The explicit form of the field equations (\ref{eq:metric_eq}) and (\ref{eq:phi_eq}) in the case of hydrostatic equilibrium can be found, for example, in Ref.~\cite{Damour1993}. 
Given a central rest-mass density value $\tilde{\rho}_c := \tilde{\rho}(0)$, the background equations can be numerically integrated subject to regularity conditions at the origin and asymptotic conditions $\nu(r) \overset{r \to \infty}{\rightarrow} 0$ and $\phi(r) \overset{r \to \infty}{\rightarrow} 0$. The stellar models thus constructed are characterized by their ADM mass $M$ and scalar charge $\omega_s$, such that $\phi = \omega_s/r + O\left(r^{-2}\right)$ asymptotically. The stellar radius $R$ is determined by $\tilde{p}(R) = 0$. 

\begin{figure}
\centering
\includegraphics[width=\linewidth]{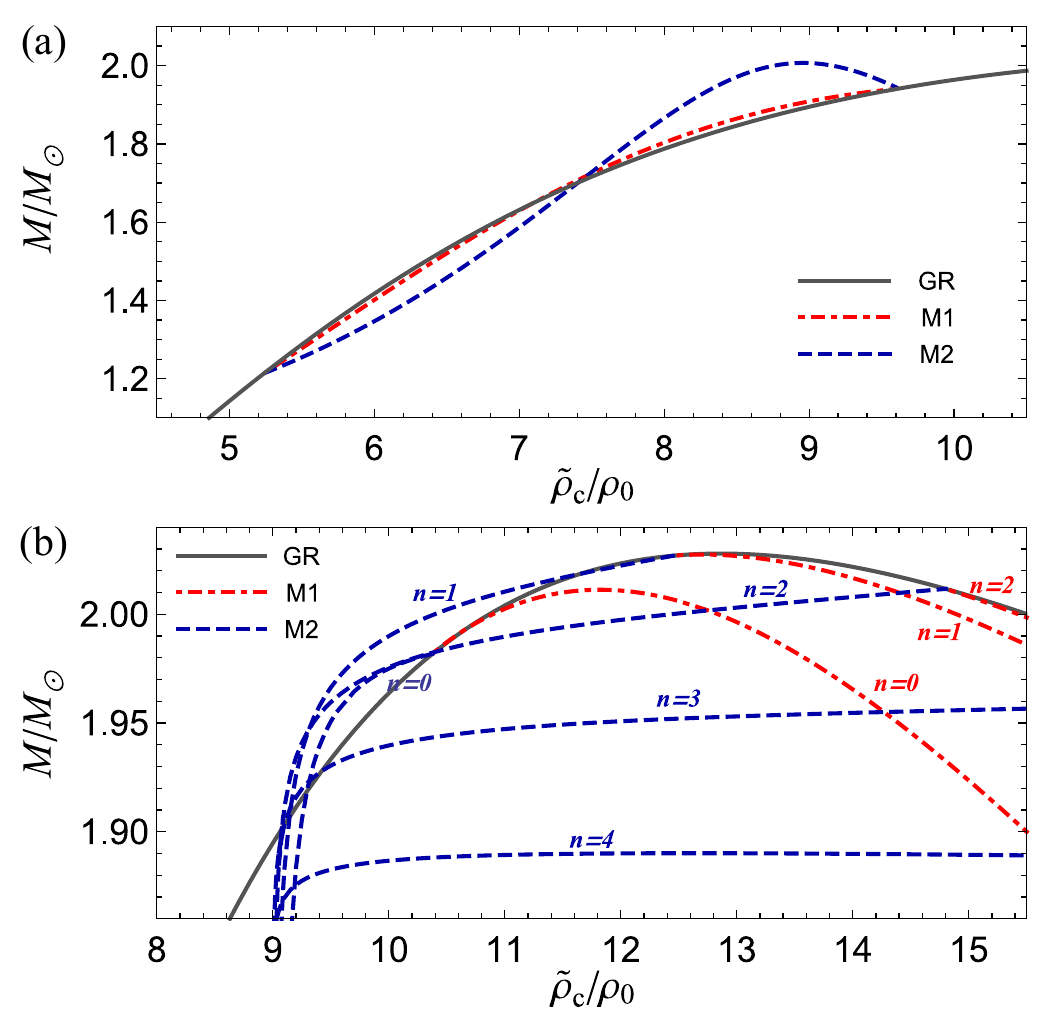}
\caption{ADM mass of sequences of equilibrium stellar configurations in M1 and M2, with (a) $\beta = -5$ and (b) $\beta=100$, parametrized by their central rest-mass density $\tilde{\rho}_c$. Branches of scalarized solutions appear at sufficiently high densities. In (b), we display only the first families of scalarized solutions, labeled by the number $n$ of nodes in their scalar field profile; see Refs.~\cite{Pani2011,Mendes2016} for details.}
\label{fig:background}
\end{figure}

Families of equilibrium solutions for Models 1 and 2 are presented in Fig.~\ref{fig:background} for $\beta =-5$ and $\beta = 100$ \footnote{The existence of scalarized solutions for $\beta>0$ is especially interesting since GR is known to be an attractor of the cosmological evolution in this class of STTs \cite{Damour1993b}, meaning that they can pass solar system tests with no need to fine tune cosmological initial conditions~\cite{Anderson2017}. This is to be contrasted with the $\beta<0$ case, where the opposite happens~\cite{Anderson2016}.}. For low enough central densities, the only equilibrium solutions have $\phi = 0$, in which case the fluid and metric functions recover their GR values. These solutions are depicted in solid gray in Fig.~\ref{fig:background}.  Nonetheless, above a certain critical central density, the scalar field can be suddenly activated: the trivial $\phi = 0$ solutions become unstable under scalar perturbations~\cite{Harada1997,Mendes2015}, and branches of ``scalarized'' configurations develop, {\it i.e.}, solutions with a nontrivial scalar field profile. For $\beta<0$, scalarized solutions in both M1 and M2 have the same qualitative behaviour, but they are radically different when $\beta > 0$ \cite{Mendes2016}. 
\\

\noindent\textit{Perturbations.---}We describe radial perturbations of the spacetime in a gauge where the metric retains form~(\ref{eq:metric}), by setting $\nu = \nu(t,r) = \nu_{(0)}(r) + \delta \nu (t,r)$ and $\lambda = \lambda(t,r) = \lambda_{(0)}(r) + \delta \lambda(t,r)$, where the subscript $(0)$ refers to background quantities. We write the perturbed scalar field as $\phi(t,r) = \phi_{(0)}(r) + \delta\phi(t,r)$. The Jordan-frame metric becomes $\tilde{g}_{\mu\nu} = \tilde{g}_{\mu\nu}^{(0)} + \tilde{h}_{\mu\nu}$, with
\begin{equation} \label{eq:htilde}
	\tilde{h}_{\mu\nu} = a_{(0)}^2 h_{\mu\nu} + 2 g_{\mu\nu}^{(0)}  a_{(0)}^2 \alpha_{(0)} \delta \phi,
\end{equation}
where $a_{(0)} := a(\phi_{(0)})$ and $\alpha_{(0)} := \alpha(\phi_{(0)})$. 

The radially perturbed configuration is characterized by a Lagrangian displacement vector field $\tilde{\xi}^\mu=(0, \xi, 0, 0)$, in terms of which we can write the Eulerian change in the fluid four-velocity $\delta \tilde{u}^\alpha=a_{(0)}^{-1} e^{-\nu_{(0)}} \left( -\delta\nu - \alpha_{(0)}\delta \phi, \partial_t \xi,0,0 \right)$. 
Under the assumption that the perturbed fluid has a negligible temperature and the same EoS as the unperturbed configuration, the pressure and energy density perturbations $\delta \tilde{p}$ and $\delta \tilde{\epsilon}$ can be related to the rest-mass density perturbation $\delta \tilde{\rho}$ through $\delta \tilde{\epsilon} = \tilde{\rho}^{-1}(\tilde{\epsilon} + \tilde{p}) \delta \tilde{\rho}$ and $\delta \tilde{p}  = \tilde{\rho}^{-1} \Gamma_1 \tilde{p} \delta \tilde{\rho}$,
where $\Gamma_1 := \partial \ln \tilde{p}/ \partial \ln \tilde{\rho}$ is the adiabatic exponent. 

Standard manipulations of the perturbed field equations [see Supplemental Material (SM)] lead to master equations for the Lagrangian displacement $\xi$ and the scalar field perturbation $\delta\phi$, in terms of which the remaining perturbation variables $\delta \lambda$, $\delta \nu$, and $\delta \tilde{\rho}$ can be determined. They read as
\begin{align} \label{eq:xi}
&e^{3\lambda_{(0)}} \! \left(\tilde{\epsilon}_{(0)} + \tilde{p}_{(0)}\right) \ddot{\xi} - \! \left[ \frac{\Gamma_1 \tilde{p}_{(0)}}{a_{(0)}^4 r^2} e^{\lambda_{(0)} + 3\nu_{(0)}} \big( e^{-\nu_{(0)}} a_{(0)}^4 r^2 \xi \big)' \right]' \nonumber \\
&+ A_\xi \xi + A_{\delta \phi} \delta \phi + A_{\delta \phi'} \delta \phi' = 0,
\end{align}
\begin{equation} \label{eq:deltaphi}
e^{2\lambda_{(0)}-2\nu_{(0)}} \delta \ddot{\phi} - \delta \phi'' + \!B_{\delta \phi'} \delta \phi' + \!B_{\delta \phi} \delta \phi + \!B_{\xi'} \xi'+ \! B_{\xi} \xi =\!0,
\end{equation}
where primes and dots indicate radial and temporal differentiation, respectively, and the $A$ and $B$ coefficients depend only on background quantities. They are displayed in the SM, together with explicit expressions for $(\delta\lambda, \delta\nu, \delta\tilde{\rho})$.

We solve the coupled second-order system (\ref{eq:xi})-(\ref{eq:deltaphi}) in the frequency domain with the ans\"atze $\xi(t,r) = \xi(r) e^{i\omega t}$ and $\delta\phi(t,r) = \delta\phi(r) e^{i\omega t}$, $\omega \in \mathbb{C}$. We require $\xi(r)$ and $\delta\phi(r)$ to be everywhere regular and $\delta\phi(t,r)$ to be purely outgoing at spatial infinity. In general, solutions satisfying these conditions exist only for a discrete set of complex frequencies $\omega$, which we aim to determine.
As a consistency check, we also solve the system (\ref{eq:xi})-(\ref{eq:deltaphi}) in the time domain and perform a Fourier analysis to unveil the vibration modes of the solution. Details on boundary conditions and numerical techniques are deferred to the SM.
\\

\noindent\textit{Radial quasinormal modes of NSs in STTs.---}Since gravity in GR is not dissipative in spherical symmetry, radial fluid oscillations are confined to the stellar interior and are described by a set of normal modes. In STTs, stellar solutions with $\phi_{(0)} = 0$ present the same spectrum of fluid oscillations as in GR: In this case, Eqs.~(\ref{eq:xi}) and (\ref{eq:deltaphi}) decouple, and Eq.~(\ref{eq:xi}) simply describes perturbations in GR.
Moreover, these GR-like stellar models display additional families of modes associated with the scalar sector.
For scalarized solutions, fluid and scalar field perturbations are coupled, and the spectrum changes accordingly. Fluid modes have their frequency shifted and gain an imaginary part, while scalar modes dictate the stability of the star. We elaborate on these ideas below for the cases of $\beta = -5$ and $\beta = 100$.

\begin{figure*}
\centering
\includegraphics[width=1.\linewidth]{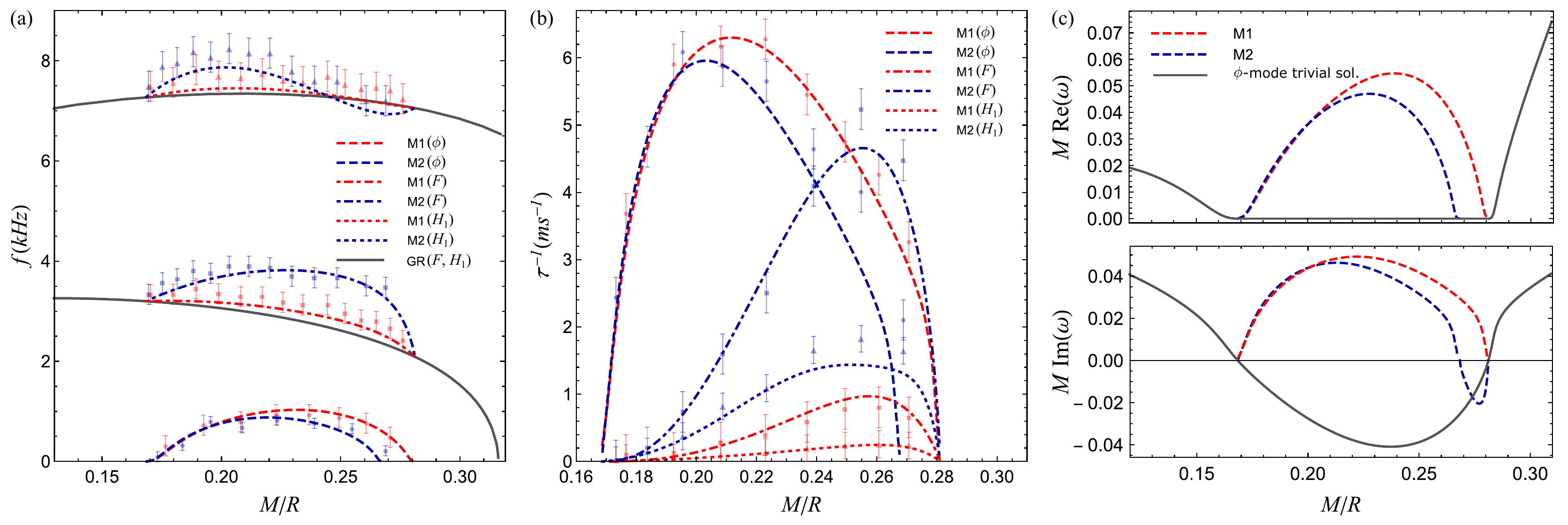}
\caption{Frequency (a) and inverse damping time (b) of the three lowest-frequency radial modes of stellar models in STTs, for M1 and M2 with $\beta = -5$, as a function of compactness. 
Lines (points) represent values computed with frequency (time) domain techniques. Panel (c) focuses on the fundamental $\phi$ mode, showing its real and imaginary parts as a function of stellar compactness. }
\label{fig:QNM_beta-5}
\end{figure*}

\begin{figure*}
\centering
\includegraphics[width=\linewidth]{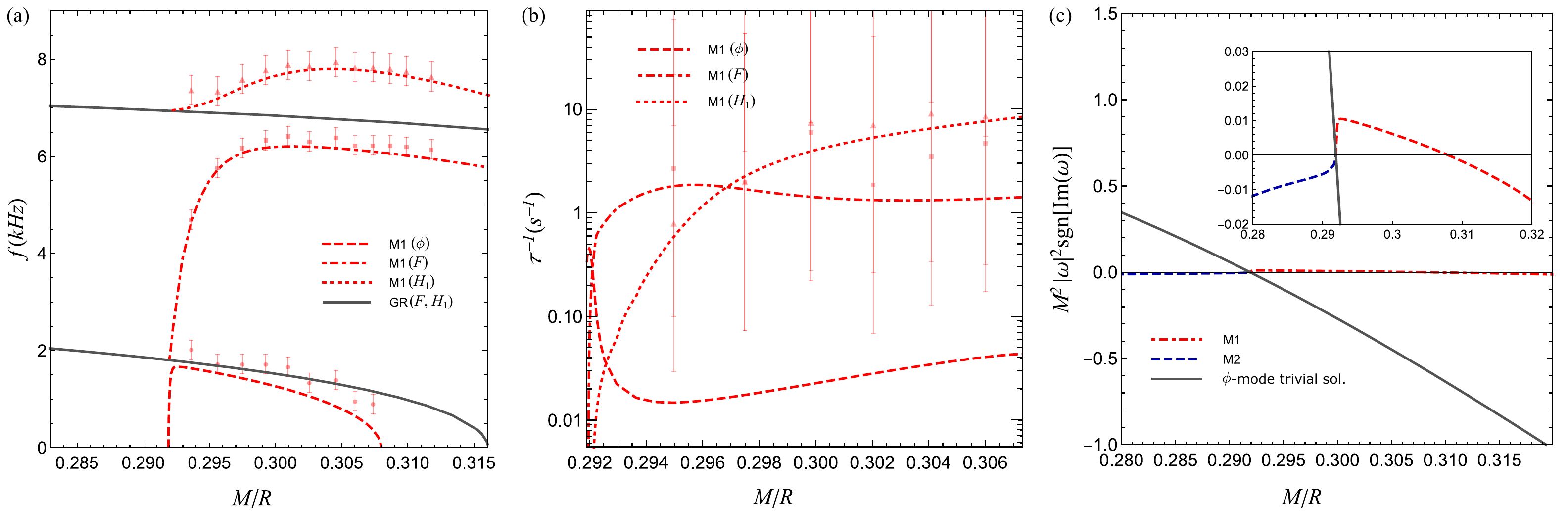}
\caption{Same as Fig.~\ref{fig:QNM_beta-5}, except for $\beta=100$. Panels (a) and (b) refer to Model 1 alone since all scalarized stars in Model 2 are unstable (see SM). For the sake of presentation, in panel (c), we plot the absolute value of the $\phi$-mode frequencies only, scaled by the sign of the imaginary part of $\omega$ to help distinguish between stable and unstable modes. Note that scalarized solutions in M2 with $n=0$ cease to exist at the same critical compactness $M/R \approx 0.292$, where scalarized solutions in M1 first appear (cf. Fig.~\ref{fig:background}\red{b}).}
\label{fig:QNM_beta100}
\end{figure*}

Figure \ref{fig:QNM_beta-5}\red{a} displays the frequency $f := \textrm{Re}(\omega)/(2\pi)$ of the first lowest-frequency radial modes for scalarized solutions in Models 1 and 2 with $\beta = -5$, as a function of stellar compactness $M/R$. The fundamental ($F$) and first overtone ($H_1$) frequencies for general relativistic stars are displayed in solid gray and have values $\lesssim 3.2$ and $\lesssim 7.3$ kHz, respectively. These modes are shifted in the presence of spontaneous scalarization, in a way analogous to that described by Sotani~\cite{Sotani2014} within the Cowling approximation---see SM for a comparison. Note that in M1, the $F$ and $H_1$ mode frequencies are less shifted than in M2, which might be traced to the fact that the scalar charge of scalarized solutions in M1 is typically smaller than in M2 \footnote{ A scalarized star with compactness $M/R = 0.21$ (which corresponds to $\tilde{\rho}_c \approx 6.56 \rho_0$ in M1 and $\tilde{\rho}_c \approx 7.05 \rho_0$ in M2) has $\omega_s/M \approx 0.17$ in M1 and $\omega_s/M \approx 0.40$ in M2.}. The inverse of the damping time $\tau := 1/\textrm{Im}(\omega)$ is shown in Fig.~\ref{fig:QNM_beta-5}\red{b} as a function of the compactness $M/R$. For all modes, it is of the order of the dynamical timescale ($\tau_\textrm{dyn}:=\sqrt{R^3/M}\sim 0.1$ms) of the star. 

Besides the QNMs that correspond to deformations of GR fluid modes, Fig.~\ref{fig:QNM_beta-5} reveals the presence of a new family of pulsation modes, which have no analogue in GR. We call them {\it scalar} or $\phi$ modes. For $\beta = -5$, they have smaller frequencies (more than 1.6 kHz lower than the $F$ mode in GR) and a slightly lower damping time than the shifted GR modes. 
Intuitively, the lower frequency of $\phi$ modes may come from the fact that scalar field oscillations are not confined to the star; thus, their wavelength is typically larger than that of fluid modes.

Figure \ref{fig:QNM_beta-5}\red{c} presents a more complete picture of the fundamental $\phi$ mode. For stellar models with $\phi_{(0)}=0$, the fundamental $\phi$ mode is shown in solid gray, while it is shown in colors for scalarized solutions. 
Before the critical compactness for spontaneous scalarization ($M/R \approx 0.168$), scalar perturbations are decoupled from fluid oscillations, and the fundamental $\phi$-mode frequency goes to zero as $M/R \approx 0.168$ is approached. For $\phi_{(0)}=0$ solutions with $0.168 \lesssim M/R \lesssim 0.281$, the fundamental $\phi$ mode becomes unstable---this is the same instability originally discussed in Ref.~\cite{Harada1997}.  As for scalarized solutions, the frequency and damping time of the fundamental $\phi$ mode were already visible in Figs.~\ref{fig:QNM_beta-5}\red{a} and~\ref{fig:QNM_beta-5}\red{b}.
Remarkably, it is the $\phi$ mode (not the $F$ mode) that becomes unstable at the turning point along a sequence of scalarized equilibrium solutions and therefore determines their (in)stability to gravitational collapse. Indeed, for M2, we see that the $\phi$-mode frequency vanishes at the compactness corresponding to the turning point in Fig.~\ref{fig:background}\red{a}, becoming unstable ($\omega^2 < 0$) thereafter. For M1 with $\beta = -5$, the sequence of scalarized solutions in Fig.~\ref{fig:background}\red{a} does not reach a local maximum before rejoining the GR branch; correspondingly, all $\phi$ modes are stable in this case.
For $M/R \gtrsim 0.281$, all equilibrium solutions again have a constant scalar field profile, and the scalar modes decouple from fluid oscillations.

In Fig.~\ref{fig:QNM_beta100}, we present the analogue of Fig.~\ref{fig:QNM_beta-5} for M1 with $\beta = 100$ (our analysis shows that all scalarized solutions for M2 with $\beta = 100$ are unstable; see SM). 
Remarkably, despite the fact that scalarized solutions in M1 with $\beta = 100$ typically have very small scalar charges \footnote{For example, a scalarized star with compactness $M/R = 0.30$ would have $\omega_s/M = 5.6 \times 10^{-5}$.}, extremely abrupt modifications to the spectrum can be seen in Fig.~\ref{fig:QNM_beta100}\red{a}. Figure~\ref{fig:QNM_beta100}\red{b} reveals that the typical damping is now much longer than the stellar dynamical timescale, reaching $\sim 100$s for the $\phi$ mode (in agreement with simulations performed in Ref.~\cite{Mendes2016}). In Fig.~\ref{fig:QNM_beta100}\red{c} it is clear how, at the critical compactness for spontaneous scalarization, the scalar mode becomes unstable along the sequence of trivial $\phi_{(0)}=0$ solutions, whereas the coupling to fluid oscillations stabilizes this mode for scalarized configurations lying before the turning point in the mass diagram in Fig.~\ref{fig:background}\red{b}.

\noindent \textit{Discussion.---}Our findings suggest that even when equilibrium stellar models in modified theories of gravity are close to GR---especially taking into account the uncertainties in the nuclear EoS---their oscillation spectrum may be radically different. In particular, new families of modes are expected to arise due to coupling to the additional degrees of freedom.

In the case of spherical oscillations in STTs, we find that the new $\phi$ modes have lower frequencies and relatively long damping times compared to GR. They could be excited in a variety of astrophysical processes. For instance, Soft Gamma Repeaters---presumably magnetars~\cite{rDcT92,cTrD95}---manifest as giant flares followed by X-ray tails with decaying times of the order of hundreds of seconds. Such tails exhibit quasiperiodic oscillations with frequencies in the range of $10^1 - 10^3$ Hz \cite{aWtS07}, which could, in principle, excite $\phi$ modes, with possible imprints in the electromagnetic emission.

Also, $\phi$ modes could be excited in binary neutron star systems. In the inspiral phase, they could become resonant with the orbital motion, draining energy from it, and thus altering the GW phase evolution \cite{Gold2012,Hinderer2016}. Because of the lower $\phi$-mode frequency, this effect could occur in STTs for larger binary separations than in GR, thus in a frequency band more suitable to current GW detectors.
In the postmerger phase, quasiradial scalar oscillations ({\it i.e.}, the modified radial $\phi$ modes in rotating systems) could be excited if a hypermassive neutron star forms \cite{tBsSmS00}. These could have nonspherical components that would emit GWs directly detectable by ground-based GW detectors.

Other astrophysical scenarios in which $\phi$ modes could be excited include gravitational collapse and phase transitions in the core of NSs.

Important ramifications of our work include studies on nonradial oscillations of spherical stars and quasiradial modes of rotating systems, extensions to other modified theories of gravity, and detectability analyses in various astrophysical scenarios.
\\

\acknowledgments
We are grateful to William East, Eric Poisson, Huan Yang, Luis Lehner, Ryan Westernacher-Schneider, Job Feldbrugge, Stephen Green, and Jonah Miller for insightful discussions and suggestions.
This work was started while R. M. was benefiting from a CITA National Fellowship at the University of Guelph. It was supported in part by the Natural Sciences and Engineering Research Council of Canada. Research at the Perimeter Institute is supported by the Government of Canada through Industry Canada and by the Province of Ontario through the Ministry of Research and Innovation.

\bibliography{../Refs/references}

\appendix
\section{\\SUPPLEMENTAL MATERIAL}
\subsection{Other perturbed quantities and coefficients of master equations}
We describe how the perturbed metric functions $\delta \lambda$ and $\delta \nu$, as well as the perturbation to the rest-mass density $\delta \tilde{\rho}$, relate to the Lagrangian displacement and scalar field perturbations, which obey Eqs.~(\ref{eq:xi}) and~(\ref{eq:deltaphi}). (See e.g.~Ref.~\cite{Friedman2013} for background on perturbation theory of relativistic fluids.)

The perturbed rest-mass density obeys 
\begin{align} \label{eq:deltan}
\delta \tilde{\rho} = & -\tilde{\rho}_{(0)} \delta \lambda - 3 \tilde{\rho}_{(0)} \alpha_{(0)} \delta \phi - \tilde{\rho}_{(0)} \xi' - \xi \tilde{\rho}_{(0)}^\prime \nonumber \\
& - \xi \tilde{\rho}_{(0)} \left( \frac{2}{r} + 3\alpha_{(0)} \Psi_{(0)} + \lambda_{(0)}' \right),
\end{align}
where $\Psi_{(0)}:= \phi'_{(0)}(r)$. This equation follows from the perturbed equation for conservation of baryon mass, $\delta[\tilde{\nabla}_\mu (\tilde{\rho} \tilde{u}^\mu)] = 0$.
The $tr$-component of the perturbed field equations can be solved for $\delta \lambda$, resulting in 
\begin{equation} \label{eq:deltalambda}
\delta \lambda =  r \Psi_{(0)} \delta \phi - 4\pi r e^{2\lambda_{(0)}} a_{(0)}^4 \left(\tilde{\epsilon}_{(0)} + \tilde{p}_{(0)}\right) \xi,
\end{equation}
while from the $rr$-component, together with Eq.~(\ref{eq:deltalambda}), one gets
\begin{align} \label{eq:deltanul}
\delta \nu' =& - 4 \pi e^{4 \lambda_{(0)}} a_{(0)}^4 \left[1 + 8 \pi r^2 a_{(0)}^4 \tilde{p}_{(0)}\right] \left(\tilde{\epsilon}_{(0)} + \tilde{p}_{(0)}\right) \xi \nonumber \\
&+ e^{2\lambda_{(0)}} \left[\Psi_{(0)} + 8\pi r \tilde{p}_{(0)} a_{(0)}^4 \left(2 \alpha_{(0)} + r\Psi_{(0)}\right)\right] \delta \phi \nonumber \\
&+ r \Psi_{(0)} \delta \phi' +4 \pi r e^{2\lambda_{(0)}} a_{(0)}^4 \delta \tilde{p}.
\end{align}

Equation (\ref{eq:xi}) is obtained from the $r$-component of the perturbed equations of motion, $\delta[\tilde{\nabla}_\nu \tilde{T}_\mu^{~\nu}] = 0$, together with Eqs.~(\ref{eq:deltalambda}) and~(\ref{eq:deltanul}), whereas Eq.~(\ref{eq:deltaphi}) stems from the perturbed scalar field equation. The $A$- and $B$-coefficients in Eqs.~(\ref{eq:xi}) and (\ref{eq:deltaphi})  are given in terms of the background quantities as follows: 
\begin{widetext}
\begin{eqnarray*}
A_\xi &=& - \frac{1}{4r^2} \left(\tilde{p}_{(0)}+ \tilde{ \epsilon}_{(0)} \right) e^{\lambda_{(0)}+2\nu_{(0)}} \left\{ e^{4 \lambda_{(0)}} + 6 e^{2 \lambda_{(0)}} - 7 + 8r \alpha_{(0)} \Psi_{(0)} \left( e^{2\lambda_{(0)}} + 1 + r^2 \Psi_{(0)}^2 \right) + 12 r^2 \Psi_{(0)}^2 \alpha_{(0)}^2
\right. \\
&+& 16 \pi r^2 e^{2 \lambda_{(0)}} a_{(0)}^4 \left[ 4 \pi e^{2 \lambda _{(0)}} \tilde{p}_{(0)}^2 r^2 a_{(0)}^4 + \tilde{p}_{(0)} \left( e^{2 \lambda_{(0)}} + 1 + 3 \alpha_{(0)}^2 + 2 r \Psi_{(0)} \alpha_{(0)} \right) - \tilde{\epsilon}_{(0)} \left(\alpha_{(0)} + r\Psi_{(0)}\right)^2\right]  \\
&+& \left.  \Psi_{(0)}^2 r^2 \left(2 e^{2 \lambda_{(0)}} + 6+ r^2 \Psi_{(0)}^2 - 4 \alpha'_{(0)} \right)  \right\} + \frac{\Gamma _1}{2 r} e^{\lambda_{(0)}+2\nu_{(0)}} \left\{ \Psi_{(0)} \tilde{p}_{(0)} \left[ r \Psi_{(0)} \left(3 e^{2 \lambda_{(0)}} + 7 - r^2 \Psi_{(0)}^2 + 2 \alpha'_{(0)} \right) \right.\right.
\\
&-& \left. \alpha_{(0)} \left( e^{2 \lambda _{(0)}} +7 - 9 r^2 \Psi_{(0)}^2 \right) -8 r \Psi_{(0)} \alpha_{(0)}^2 \right] - \Psi_{(0)} \tilde{\epsilon}_{(0)}  \left( \alpha_{(0)} - r \Psi_{(0)}\right) \left(e^{2\lambda_{(0)}} - 1 + 2 r \alpha_{(0)} \Psi_{(0)} + r^2 \Psi_{(0)}^2 \right)
\\
&-&\left. 8 \pi r e^{2 \lambda_{(0)}} \tilde{p}_{(0)} a_{(0)}^4 \left[ \tilde{p}_{(0)} \alpha_{(0)} \left (3 \alpha_{(0)} - 7 r \Psi _{(0)} \right) - \tilde{\epsilon}_{(0)} \left(\alpha_{(0)} - 2 r \Psi_{(0)}\right)  \left(\alpha_{(0)} + r\Psi_{(0)}\right)\right] \right\} \\
&+& \tilde{p}_{(0)} \Psi_{(0)} \Gamma_1' e^{\lambda_{(0)}+2\nu_{(0)}} \left(\alpha_{(0)} - r \Psi_{(0)}\right),\\
A_{\delta\phi} &=& \frac{1}{2r} \left(\tilde{p}_{(0)}+\tilde{\epsilon}_{(0)}\right) e^{\lambda_{(0)}+2\nu_{(0)}} \left[ 8 \pi r^2 \tilde{p}_{(0)} e^{2 \lambda_{(0)}} a_{(0)}^4 \left( \alpha_{(0)} + r \Psi_{(0)} \right) - \alpha_{(0)} \left( 3 e^{2 \lambda_{(0)}} - 3 + 5 r^2 \Psi_{(0)}^2 \right) - 6 r \Psi_{(0)}\alpha_{(0)}^2 \right.
\\
&+& \left. r \Psi_{(0)} \left( e^{2 \lambda_{(0)}} + 1 - r^2 \Psi_{(0)}^2 + 2 \alpha'_{(0)}\right) \right]
+ \frac{\Gamma _1}{2 r} e^{\lambda_{(0)}+2\nu_{(0)}} \left[ - 8 \pi r^2 \tilde{p}_{(0)} \tilde{\epsilon}_{(0)} e^{2 \lambda_{(0)}} a_{(0)}^4 \left( \alpha_{(0)} + r \Psi_{(0)}\right) \right.\\
&+& \left. \tilde{\epsilon}_{(0)} \left( e^{2 \lambda_{(0)}} - 1 \right) \left( 3\alpha_{(0)} + r \Psi_{(0)}\right)
+r \tilde{\epsilon}_{(0)} \Psi_{(0)} \left( 5 r  \alpha_{(0)} \Psi_{(0)} + 6 \alpha_{(0)}^2 + r^2 \Psi_{(0)}^2 \right)
+ 2 r \tilde{p}_{(0)} \Psi_{(0)} \left( e^{2 \lambda_{(0)}} - 3 \alpha'_{(0)} \right) \right] \\
&-& \tilde{p}_{(0)} \Gamma_1' e^{\lambda_{(0)}+2\nu_{(0)}} \left( 3 \alpha_{(0)} + r \Psi_{(0)} \right),\\
A_{\delta\phi'} &=& e^{\lambda_{(0)}+2\nu_{(0)}} \left[\left( \tilde{p}_{(0)}+ \tilde{\epsilon}_{(0)} \right) \left( \alpha_{(0)} + r \Psi_{(0)}\right) - \Gamma_1 \tilde{p}_{(0)} \left(3 \alpha_{(0)} + r \Psi_{(0)}\right) \right],
\end{eqnarray*}
\begin{eqnarray*}
B_{\delta \phi'} &=& - \frac{1}{r} \left( e^{2\lambda_{(0)}} + 1 \right) + 4 \pi r e^{2\lambda_{(0)}} a_{(0)}^4 \left(\tilde{\epsilon}_{(0)}-\tilde{p}_{(0)}\right) , \\
B_{\delta \phi} &=& 4 \pi a_{(0)}^4 \tilde{p}_{(0)} e^{2 \lambda_{(0)}} \left\{ \left( 3 \alpha_{(0)} + r \Psi_{(0)}\right ) \left[ \alpha_{(0)} (3 \Gamma_1 - 5) + r \Psi_{(0)}(\Gamma_1-3) \right] - 3\alpha'_{(0)}\right\} \\
&+& 4\pi a_{(0)}^4 \tilde{\epsilon}_{(0)} e^{2 \lambda_{(0)}} \left[ \left( \alpha_{(0)}+ r \Psi_{(0)}\right)^2+\alpha'_{(0)}\right] - 2 e^{2 \lambda_{(0)}}\Psi _{(0)}^2,\\
B_{\xi'} &=& 4 \pi  a_{(0)}^4 e^{2 \lambda_{(0)}} \left\{ \tilde{p}_{(0)} \left[ (3 \Gamma_1-1)\alpha_{(0)} + (\Gamma_1-1) r \Psi_{(0)} \right] - \tilde{\epsilon}_{(0)} \left( \alpha_{(0)} + r \Psi_{(0)}\right) \right\}, \\
B_\xi &=& -\frac{2 \pi  e^{2 \lambda _{(0)}} a_{(0)}^4}{\Gamma_1 r \tilde{p}_{(0)}} \left\{ - \left( \tilde{p}_{(0)} + \tilde{\epsilon }_{(0)}\right)^2 \left( \alpha_{(0)} + r \Psi_{(0)} \right) \left( 8 \pi r^2 a_{(0)}^4 e^{2 \lambda_{(0)}} \tilde{p}_{(0)} + e^{2 \lambda_{(0)}} -1 + 2r \alpha_{(0)} \Psi_{(0)} + r^2 \Psi_{(0)}^2 \right)
 \right.
\\
&+& \tilde{p}_{(0)}^2 \Gamma_1^2 (3 \alpha_{(0)} + r \Psi_{(0)}) \left(8 \pi r^2 a_{(0)}^4 e^{2 \lambda_{(0)}} \tilde{p}_{(0)} + e^{2 \lambda_{(0)}} - 5 - 6 r \alpha_{(0)} \Psi_{(0)} - r^2 \Psi_{(0)}^2 \right) - 2 \tilde{p}_{(0)} \Gamma_1 \left( \tilde{p}_{(0)} + \tilde{\epsilon }_{(0)}\right) 
\\
&\times & \left[ \alpha_{(0)} \left( 8 \pi r^2 a_{(0)}^4 e^{2 \lambda_{(0)}} \left( 2 \tilde{p}_{(0)} - \tilde{\epsilon }_{(0)} \right) - e^{2 \lambda_{(0)}} -1 \right) - 6 r \alpha_{(0)} \Psi_{(0)} \left( \alpha_{(0)} + r\Psi_{(0)} \right) \right.
\\
&+& \left. \left.
r\Psi_{(0)} \left( 8 \pi r^2 a_{(0)}^4 e^{2 \lambda_{(0)}} \left( \tilde{p}_{(0)} - \tilde{\epsilon }_{(0)} \right) + 2 e^{2 \lambda_{(0)}} - 2 -r^2 \Psi_{(0)}^2 \right) \right] \right\}.
\end{eqnarray*}
\end{widetext}

The Lagrangian displacement $\xi$ is only defined in the stellar interior ($r\leq R$). Outside the star ($r>R$), the only nonvanishing coefficients in Eq.~(\ref{eq:deltaphi}) are
\begin{equation*}
B_{\delta \phi'} = -\frac{1+e^{2\lambda_{(0)}}}{r}, 
\quad
B_{\delta \phi} = -~2 e^{2\lambda_{(0)}} \Psi_{(0)}^2.
\end{equation*}

\subsection{Unstable modes of scalarized solutions with $\beta>0$}
\label{sec:unstable}
Here we analyze the stability of scalarized solutions in STTs with $\beta>0$. For theories with $\beta <0$, and in particular for the model $a(\phi)=\exp(\beta \phi^2/2)$ (our Model 2) popularized by the works of Damour and Esposito-Far\`ese, stability of scalarized solutions was investigated by Harada~\cite{Harada1998} through a turning point technique.
In Model 2 with $\beta>0$, however, turning point techniques do not seem to be directly applicable. As a by-product of our work, here we examine instability to collapse for scalarized solutions in M1 and M2 with $\beta = 100$ (but our results seem generic for any $\beta>0$).

To interpret our results it is useful to recall our thermodynamic expectations. It is well known that in GR a sequence of stellar models is secularly unstable on one side of an extremum in a mass diagram~\cite{Sorkin1982}. For spherical models, a turning point also signals the onset of dynamical instability, when a radial mode changes its frequency from real to imaginary. It is easy to see how dynamical and secular stabilities are related in this case. The point of marginal stability is characterized by the existence of a zero-frequency mode, which is just a time-independent perturbation linking two nearby equilibrium solutions. From the first law of thermodynamics, a perturbation that keeps the star in equilibrium satisfies $\delta M = \mathcal{E} \delta M_b$ where $\mathcal{E}$ is the injection energy and $M_b$ is the baryon mass. Thus, for a zero frequency perturbation involving no change in baryon mass we have $\delta M = 0$, which is the necessary condition for an extremum in a mass diagram. The same argument holds in STTs, as discussed by Harada~\cite{Harada1998}. The only difference is that now the control space of spherically symmetric equilibrium models is not only the baryon number, but also the asymptotic value of the scalar field, $\phi_0$, so that the first law reads $\delta M~=~\mathcal{E} \delta M_b - \omega_s \delta \phi_0$, where $\omega_s$ is the scalar charge. Thus $\delta M =0$ for a zero frequency perturbation that involves no change in $M_b$ or $\phi_0$.

\begin{figure}
\includegraphics[width=\linewidth]{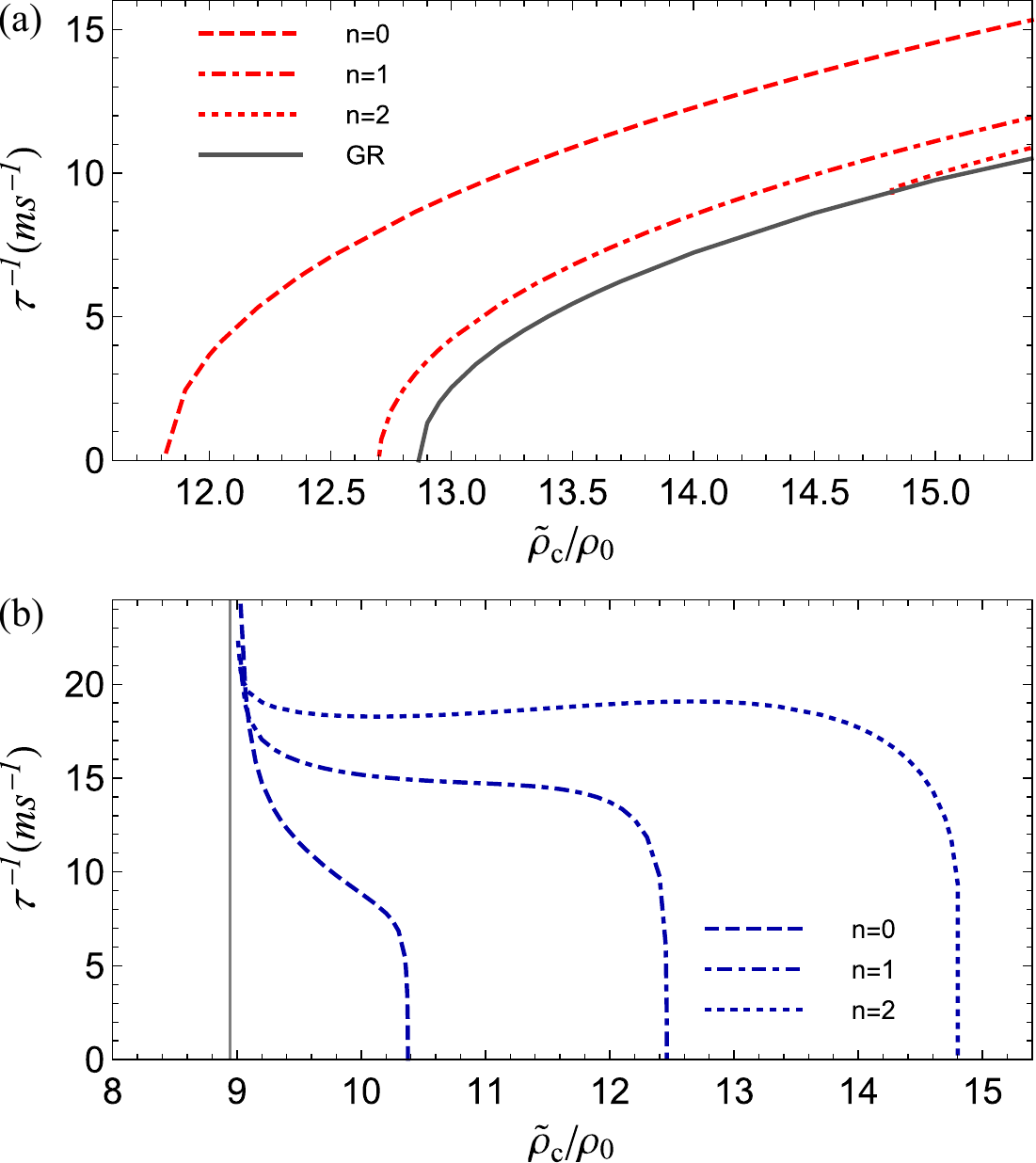}
\caption{Inverse of the instability timescale $\tau$ as a function of the central rest-mass density of unstable stars, for (a) M1 and (b) M2 with $\beta = 100$.}
\label{fig:unstable}
\end{figure}

Our results conform to these thermodynamic considerations. Figure~\ref{fig:unstable}\red{a} shows the inverse of the instability timescale $\tau$ (such that $\delta \phi, \xi \propto e^{t/\tau}$), as a function of the normalized central rest-mass density $\tilde{\rho}_c/\rho_0$ for M1 with $\beta = 100$. We see that the branch of scalarized solutions with $n=0$ displays unstable modes for central densities $\tilde{\rho}_c\gtrsim 11.82 \rho_0$, which is consistent with the turning point in Fig.~\ref{fig:background}\red{b}. The same applies for the family of scalarized solutions with $n=1$. As can be seen back in Fig.~\ref{fig:background}\red{b}, these solutions exist for stellar models with central densities $\tilde{\rho}_c \gtrsim 12.46\rho_0$, and closer inspection reveals that the turning point in the mass diagram occurs at $\tilde{\rho}_c \approx 12.70 \rho_0$. In agreement with thermodynamic expectations, in Fig.~\ref{fig:unstable}\red{a} we see that, in the $n=1$ case, unstable modes only exist for $\tilde{\rho}_c \gtrsim 12.70 \rho_0$. Therefore, there is only a small range of densities for which scalarized solutions with $n=1$ have no unstable radial modes. At around $\tilde{\rho}_c \approx 14.80 \rho_0$ a third branch of scalarized solutions develop. From Fig.~\ref{fig:unstable}\red{a} we see that they all are unstable. This is also the case for families of solutions with higher values of $n$. In all cases, the instability timescale roughly decreases as $1/\sqrt{\tilde{\rho}_c}$ as the central density increases.

With respect to Model 2, in Ref.~\cite{Mendes2016} we found no evidence of the existence of stable scalarized solutions with $\beta>0$, so we conjectured that all such scalarized solutions were unstable. This conjecture is confirmed by the perturbative analysis in this work. In Fig.~\ref{fig:unstable}\red{b} we show $\tau^{-1}$ for scalarized stars in M2 with $\beta = 100$, for the three lowest-$n$ branches. We see that as soon as scalarized solutions appear, they quickly develop unstable radial modes. As the density approaches $\tilde{\rho}_c \approx 8.945 \rho_0$ from the right, the instability timescale goes to zero---the pattern repeats for higher $n$ families. This is the critical density at which the trace of the energy-momentum tensor vanishes in the stelar center; below this density there are no scalarized solutions \cite{Mendes2016}. 

\subsection{Radial eigenfunctions}
\begin{figure}
\includegraphics[width=\linewidth]{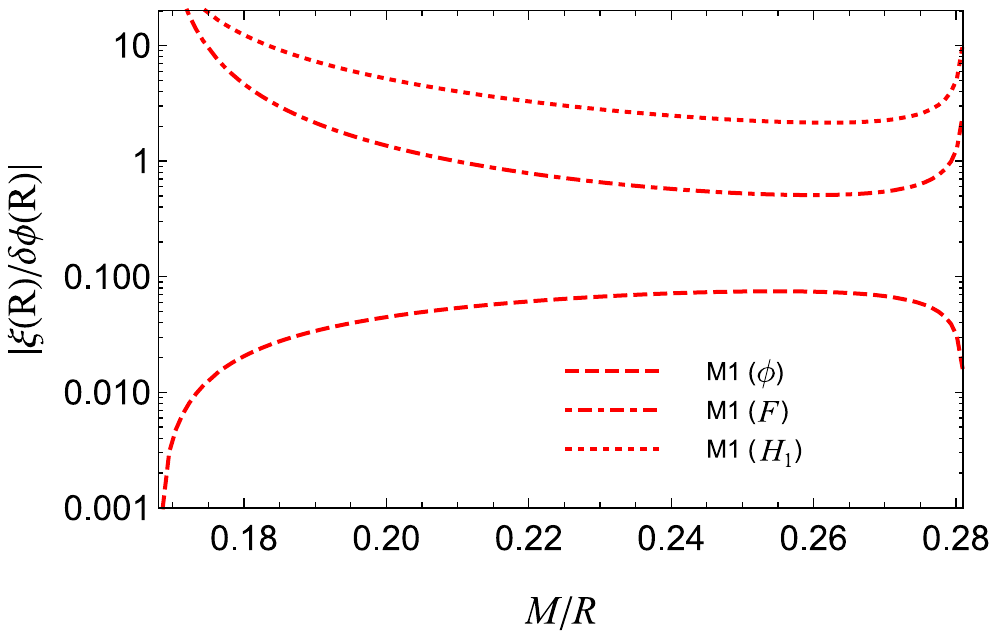}
\caption{Comparison between radial eigenfunctions of the Lagrangian displacement and scalar field perturbation at the stellar surface, for the fundamental $\phi$-mode and the shifted $F$ and $H_1$ modes, as a function of the stellar compactness, in M1 with $\beta = -5$.}
\label{fig:xitophi}
\end{figure}

In the main text, we interpreted the lowest frequency oscillation mode of scalarized stars as a deformation of a pure scalar mode of $\phi_{(0)}=0$ configurations. Correspondingly, overtones of scalarized stars were interpreted as deformations of pure fluid modes of those trivial solutions. Here we give further support for this interpretation.
As an example, we focus on scalarized solutions in Model 1 with $\beta=-5$. The lower the scalar charge---{\it i.e.} the closer the compactness to the merging points with the branch of GR solutions, which occur at $M/R \approx 0.168$ and $M/R \approx 0.281$---, the lowest-frequency mode is expected to be predominantly a scalar perturbation, while the overtones are expected to be mainly fluid perturbations. This expectation is verified in Fig.~\ref{fig:xitophi}, where we compare the radial eigenfunctions of the Lagrangian displacement $\xi(r)$ and the scalar field perturbation $\delta\phi(r)$ at the surface of the star ($r=R$), for a sequence of scalarized solutions. For the lowest frequency mode, we see that the ratio $|\xi(R)/\delta\phi(R)|$ goes to zero as the transition to GR solutions is approached, while it diverges for the overtone modes.

\subsection{Comparison to the Cowling approximation}
In GR, the Cowling approximation consists in keeping the metric fixed and only allowing for fluid perturbations (see for instance Ref.~\cite{Friedman2013}). It becomes increasingly accurate for fluid perturbations that are confined to a small region inside the star or have a short wavelength~\cite{Lindblom1990,Yoshida1997,Yoshida2001}. This is the case, for instance, of torsional oscillations confined to the stellar crust and high multipole QNMs.
The rationale behind it can be easily understood in the Newtonian setting: If $\lambda$ is the characteristic wavelength of the mode, then the Laplacian of the gravitational potential $\Phi$ is roughly $\nabla^2 \Phi \sim \Phi/\lambda^2$, and Poisson's equation implies that $\Phi \sim \rho \lambda^2$, which goes to zero as $\lambda \to 0$.

In STTs, the presence of a scalar field as an additional mediator of the gravitational interaction introduces an extra dilemma of whether or not to keep the scalar field fixed in the Cowling approximation scheme. Indeed, previous works made different choices, either keeping both the metric and the scalar field fixed (as in Ref.~\cite{Sotani2004,Yazadjiev2017}), or allowing for scalar field perturbations while keeping the Jordan-frame metric fixed (as in Refs.~\cite{Silva2014,Sotani2014}). 
Here we compare our results with those obtained with the latter procedure, which was employed in Ref.~\cite{Sotani2014} in the study of radial oscillations of spherically symmetric scalarized stars. In this case, one sets $\tilde{h}_{\mu\nu} = 0$, and Eq.~(\ref{eq:htilde}) implies that $h_{\mu\nu} = -2 g_{\mu\nu}^{(0)} \alpha_{(0)} \delta \phi$. Manipulations of the perturbed field equations allow all perturbation variables to be written in terms of the Lagrangian displacement, which follows the master equation (3.15) of Ref.~\cite{Sotani2014}. (Strictly speaking, this equation is only valid for M2, but it suffices to substitute the combination $\beta \phi $ by $\alpha(\phi)$ and the remaining $\beta$'s by $d\alpha/d\phi$ to obtain the general expression.) This version of the Cowling approximation allows scalar waves to be sourced by fluid perturbations (see Eq.~(3.12) of Ref.~\cite{Sotani2014}), although they do not cause damping to the fluid motion, which is described by a set of normal modes.

\begin{figure}
\centering
\includegraphics[width=\linewidth]{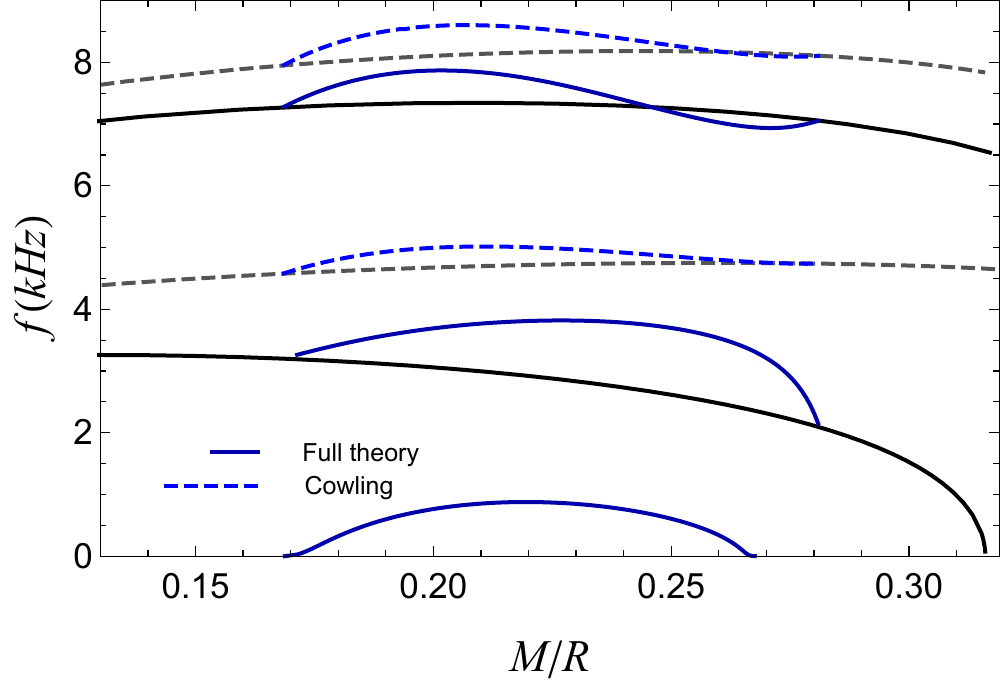}
\caption{Frequency of the first radial modes of stellar models in STTs, using full perturbation theory (solid lines) and using the Cowling approximation (dashed lines), for Model 2 with $\beta = -5$, as a function of stellar compactness.}
\label{fig:cowling}
\end{figure}

Figure~\ref{fig:cowling} shows a comparison between the mode frequencies obtained with our full analysis and with the Cowling approximation for M2 with $\beta = -5$. We see that the Cowling approximation overestimates the mode frequencies, especially for the fundamental fluid mode, and does not properly capture information about instability to collapse. More importantly, it does not reveal the presence of the additional families of modes reported in this work. Therefore, although the Cowling approximation can yield relatively accurate results, especially for higher overtones and higher angular momentum modes, it may fail to capture key aspects of stellar QNMs in modified theories of gravity.

\subsection{Frequency domain approach}
Here we describe the frequency domain techniques that we implement in order to solve the perturbation system~(\ref{eq:xi})-(\ref{eq:deltaphi}), with the oscillatory ans\"atze
\begin{equation}\label{eq:modes}
\xi(t,r) = \xi(r) e^{i\omega t}, \quad \delta \phi (t,r) = \delta\phi (r) e^{i\omega t}, \quad \omega \in \mathbb{C}.
\end{equation}

\subsubsection{Boundary conditions}

Appropriate boundary conditions for the solutions are the following.
\begin{itemize}
\item[(i)] The radial eigenfunctions $\xi(r)$ and $\delta\phi (r)$ are required to be analytic around $r=0$. Equations (\ref{eq:xi}) and (\ref{eq:deltaphi}) imply that, in a neighbourhood of $r = 0$, 
\begin{equation*}
\xi(r) = \sum_{i=0}^N a_{2i+1} r^{2i+1}, \qquad
\delta \phi(r) = \sum_{i=0}^N b_{2i} r^{2i},
\end{equation*}
for some $N\in \mathbb{Z}^+$, where $(a_1, b_0)$ are free constants. The remaining coefficients are determined recursively in terms of $(a_1, b_0)$ and the background quantities.

\item[(ii)] At the perturbed stellar surface, the pressure vanishes, implying that its Lagrangian variation is zero, $\Delta \tilde{p}|_{r=R} = 0$. But since
\begin{align*}
\Delta \tilde{p} =& -\Gamma_1 \tilde{p}_{(0)} \left\{ r \Psi_{(0)} \delta \phi + \xi' \right. \nonumber \\
& \left. +~\xi \left[3 \alpha_{(0)} \Psi_{(0)} - \nu'_{(0)} + r \Psi_{(0)}^2 + 2/r \right] \right\},
\end{align*}
and $\tilde{p}_{(0)}(R) = 0$, this condition is automatically satisfied as long as the term within brackets is regular at the stellar surface. Therefore, it is enough to impose regularity of $\xi$ and $\delta \phi$ at $r=R$.
Dividing Eq.~(\ref{eq:xi}) by $\tilde{p}_{(0)}$, and taking into account our choice of EoS in the stellar crust, we see that the terms which are potentially diverging in the limit $r\to R$ are those proportional to $\tilde{\epsilon}_{(0)}/\tilde{p}_{(0)} \propto \tilde{\rho}_{(0)}^{(1-\gamma)} \propto (r-R)^{-1}$. Therefore, we require the coefficient of the $\tilde{\epsilon}_{(0)}/\tilde{p}_{(0)}$ term to vanish at $r=R$, which amounts to requiring that the quantity
\begin{align}\label{eq:f}
&f(r) := \gamma_2 \xi' \left[e^{2 \lambda_{(0)}} -1 + r \Psi_{(0)} \left( 2 \alpha_{(0)} + r\Psi_{(0)} \right)\right]\nonumber \\
&+ \! 2 r \delta \phi' \left( \alpha_{(0)} + r \Psi_{(0)} \right)
\! +\!  \frac{\xi}{2r} \Big\{\!\! -4 e^{2\lambda_{(0)} - 2 \nu_{(0)}} r^2 \omega^2 \!- 5 \gamma_2
\nonumber \\
&+  \! 6 (\gamma_2 - 1) \left(e^{2\lambda_{(0)}} + 2 r^2 \alpha_{(0)}^2 \Psi_{(0)}^2 \right) - e^{4\lambda_{(0)}} (\gamma_2 + 1) \!+ 7
\nonumber \\
& +\! 4 r\alpha_{(0)}\! \Psi_{(0)} \! \left[ \left(1+e^{2\lambda_{(0)}}\right)(\gamma_2-2) + 2r^2 (\gamma_2-1)\Psi_{(0)}^2 \right]
\nonumber \\
& + \!r^2 \Psi_{(0)}^2 \! \left[  4 \alpha'_{(0)} - 6 - 2 e^{2\lambda_{(0)}} + 4\gamma_2 + r^2 (\gamma_2\!-\!1) \Psi_{(0)}^2\right]\!\!\Big\}
\nonumber \\
& +\! \delta \phi \Big\{ (\gamma_2 - 1) \alpha_{(0)} \left(3 e^{2\lambda_{(0)}} - 3 + 5 r^2 \Psi_{(0)}^2\right)
\nonumber \\
& +\! r\Psi_{(0)} \left[ e^{\lambda_{(0)}}(\gamma_2 + 1) + 2 \alpha'_{(0)} + (\gamma_2 - 1) ( r^2 \Psi_{(0)}^2 - 1) \right]
\nonumber \\
& + \!6 r (\gamma_2-1) \alpha_{(0)}^2 \Psi_{(0)}\Big\}
\end{align}
vanishes at the stellar surface, {\it i.e.} $f(R)=0$.

\item[(iii)] We impose a purely outgoing boundary condition for the scalar field perturbation at spatial infinity,
\begin{equation} \label{eq:outgoing}
\lim_{r \to \infty}\delta\phi(t,r) \propto e^{i \omega (t-r)}.
\end{equation}
To write this more precisely, we first notice that setting $\delta \phi (r) = Z(r)/r$, Eq.~(\ref{eq:deltaphi}) for $r>R$ can be cast in the alternative form
\begin{equation} \label{eq:Z}
\frac{d^2 Z}{dr_*^2} + \left[ \omega^2 + e^{2\nu_{(0)}} \left( 2 \Psi_{(0)}^2 + \frac{e^{-2\lambda_{(0)}} - 1}{r^2} \right) \right] Z = 0,
\end{equation}
where the tortoise coordinate $r_*$ is defined through $dr_* = e^{\lambda_{(0)} -\nu_{(0)}} dr$.
Equation~(\ref{eq:Z}) admits two linearly independent solutions, $Z^{\pm}(r)$, with asymptotic expansions
\begin{equation} \label{eq:Zpm}
Z^{\pm} (r\to \infty) = e^{\pm i\omega r_*} \sum_{i=0}^N \frac{c_i^\pm}{r^i},
\end{equation}
where $N\in \mathbb{Z}^+$ and $c_0^\pm$ are specifiable constants, in terms of which the coefficients $c_i^\pm$, $i\geq 1$, can be determined by solving Eq.~(\ref{eq:Z}) order by order. The purely outgoing boundary condition singles out $Z^-(r)$ as the physical solution. 
\end{itemize}

In summary, our task is to solve the coupled system (\ref{eq:xi})-(\ref{eq:deltaphi}) subject to a set of {\it four} boundary conditions: regularity at $r=0$ [specifically, $\xi(0) = 0$ and $\delta \phi' (0) = 0$], regularity at $r=R$ [specifically, $f(R) =0$ with $f$ defined in Eq.~(\ref{eq:f})], and an outgoing boundary condition for the scalar perturbation at spatial infinity. However, since the perturbation equations (\ref{eq:xi})-(\ref{eq:deltaphi}) are homogeneous, there is an additional overall normalization freedom to the perturbed quantities. This means that the boundary value problem posed above is overdetermined. Therefore, solutions exist only for a discrete set of values of $\omega$. Below we detail the numerical techniques that we employ to search for these frequencies.

\subsubsection{Solution to the inner problem}\label{sec:inner}
In the domain $r\in [0,R]$, the perturbation equations (\ref{eq:xi}) and (\ref{eq:deltaphi}) with the ansatz~(\ref{eq:modes}) are to be solved subject to the boundary conditions $\xi(0) = 0$, $\delta\phi'(0) = 0$, and $f(R) = 0$. To fix the normalization freedom we require $\delta \phi(R) = 1$. 

Our strategy relies on the fact that for a homogeneous system of the form 
\begin{equation} \label{eq:matrix_eq}
\frac{d {\bf x}(t)}{dt} = {\bf M}(t) {\bf x}(t),
\end{equation}
where ${\bf x}(t)$ is a vector function and ${\bf M}(t)$ is a $n \times n$ matrix function, every solution can be expressed as
\begin{equation*}
{\bf x}(t) = c_1 {\bf x}_1(t) + \cdots + c_n {\bf x}_n(t),
\end{equation*}
where $c_i$ are constants and $\{{\bf x}_1, {\bf x}_2,..., {\bf x}_n\}$ is a set of linearly independent solutions of the system (see, {\it e.g.} Ref.~\cite{Lindblom1983}). It is straightforward to rewrite the perturbation equations (\ref{eq:xi}) and (\ref{eq:deltaphi}) for the state vector ${\bf x}(r) = (\xi,\xi',\delta\phi,\delta\phi')$ in the form (\ref{eq:matrix_eq}). At $r=0$, we select two linearly independent vectors ${\bf x}(0)$ satisfying the boundary conditions $\xi(0)=0$ and $\delta \phi'(0)=0$ [say, $(0,1,1,0)$ and $(0,1,-1,0)$] and numerically integrate the differential equations from $r=0$ to $r=r_t$ (where $r=r_t$ is the transition point of the two polytropic phases, $\tilde{\rho}(r_t) = \tilde{\rho}_t$) with such initial conditions. The integration yields two solutions, ${\bf x}_1(r)$ and ${\bf x}_2(r)$ defined in the domain $0\leq r \leq r_t$ and satisfying the correct boundary conditions at $r=0$. The unique solution that satisfies all the boundary conditions must be a linear combination of the form ${\bf x}(r) = a_1 {\bf x}_1(r) + a_2 {\bf x}_2(r)$ for $0\leq r \leq r_t$, and some constants $a_i$, $i\in\{1,2\}$. We repeat this procedure for $r_t \leq r \leq R$. Namely, we select three independent vectors ${\bf x}(R)$ that satisfy the boundary condition $f(R)=0$ at $r=R$ [say, $(1,\xi'(R),1,-1)$, $(1,\xi'(R),-1,1)$, and $(1,\xi'(R),1,1)$, with $\xi'(R)$ determined in each case by solving $f(R)=0$] and numerically integrate the system from $r=R$ to $r=r_t$. This yields three solutions, ${\bf x}_3(r)$, ${\bf x}_4(r)$, and ${\bf x}_5(r)$, defined in the domain $r_t\leq r \leq R$, which satisfy the correct boundary condition at $r=R$. Again, the unique solution that satisfies all the boundary conditions must be a linear combination of the form ${\bf x}(r) = a_3 {\bf x}_3(r) + a_4 {\bf x}_4(r)+ a_5 {\bf x}_5(r)$ for $r_t\leq r \leq R$ and some constants $a_i$, $i\in\{3,4,5\}$. Finally, we require that ${\bf x}(r)$ satisfies the proper (dis)continuity condition at $r_t$, namely,
\begin{align} \label{eq:continuity}
0 &= (0,\Delta, 0, 0) +(1,\gamma_1,1,1) . [a_1 {\bf x}_1(r_t) + a_2 {\bf x}_2(r_t)] \nonumber \\
&- (1,\gamma_2,1,1). [a_3 {\bf x}_3(r_t) + a_4 {\bf x}_4(r_t) + a_5 {\bf x}_5(r_t)],
\end{align}
where 
\begin{align} \label{eq:Delta}
\Delta & := (\gamma_2 - \gamma_1) \left[  \left(\nu'_{(0)} -3 \alpha_{(0)}\Psi_{(0)} - r \Psi_{(0)}^2 -\frac{2}{r} \right) \xi \right. \nonumber \\
& \left. - \left( 3 \alpha_{(0)} + r \Psi_{(0)} \right) \delta\phi \right]_{r_t},
\end{align}
as will be derived below. Equation (\ref{eq:continuity}) gives a set of four algebraic conditions that must be satisfied by the five $a_i$'s. The overall normalization of the solution closes the algebraic system. For instance, with the choices made above, the normalization $\delta \phi (R)= 1$ imposes $a_3 -a_4 + a_5 = 1$.

Finally, we explain the origin of the $\Delta$ term in Eq.~(\ref{eq:Delta}). For a two piece polytrope, the adiabatic index can be written as $\Gamma_1 = \gamma_1 - (\gamma_1 - \gamma_2) \Theta(r-r_t)$, where $\Theta$ denotes the Heaviside step function. As a consequence, the radial derivative of $\Gamma_1$, which appears in Eq.~(\ref{eq:xi}), is singular at $r=r_t$: $\Gamma_1' = (\gamma_2 - \gamma_1) \delta (r-r_t)$. This implies that $\xi'$ will display a jump discontinuity across $r=r_t$. To obtain the magnitude of the discontinuity, we  integrate Eq.~(\ref{eq:xi}) from $r^-_t := r_t-\epsilon$ to $r^+_t := r_t+\epsilon$, taking the limit $\epsilon \to 0$. Being explicit,
\begin{align*}
&\lim_{\epsilon \to 0} \left[ \Gamma_1 \left( -\nu'_{(0)} \xi + 4 \alpha_{(0)} \Psi_{(0)} \xi + \frac{2}{r} \xi + \xi'\right)\right]_{r_t^-}^{r_t^+} \\ 
&=\! (\gamma_2 - \gamma_1) \left[ \Psi_{(0)}\! \left(\alpha_{(0)} - r \Psi_{(0)}\right) \xi
 - \! \left( 3 \alpha_{(0)} + r \Psi_{(0)} \right) \delta\phi \right]_{r_t}\!\!.
\end{align*}
Therefore, the magnitude of the jump discontinuity is
\begin{equation}
\gamma_2 \xi'(r_t^+) - \gamma_1 \xi'(r_t^-) = \Delta,
\end{equation}
where $\Delta$ is given in Eq.~(\ref{eq:Delta}).

\subsubsection{Solution to the outer problem}
For $r>R$, our task is to solve the differential equation
\begin{equation} \label{eq:deltaphioutmode}
\delta \phi'' + \frac{1+e^{2\lambda_{(0)}}}{r} \delta \phi' + e^{2\lambda_{(0)}} \left( 2 \Psi_{(0)}^2 + e^{- 2\nu_{(0)}} \omega^2 \right) \delta \phi = 0,
\end{equation}
with the purely outgoing boundary condition at spatial infinity
\begin{equation}
\lim_{r \to \infty} \delta \phi \propto \frac{ e^{-i \omega r_*}}{r}.
\end{equation}
Alternatively we can consider Eq.~(\ref{eq:Z}) for $Z(r) = r \delta \phi (r)$, subject to the condition that the solution asymptotes to $Z^-(r)$.

For an unstable mode, $\textrm{Im}(\omega) < 0$, and $\delta \phi(r) \to 0$ as $r \to \infty$. Note that with this condition the boundary value problem is Hermitian, thus the eigenfrequencies $\omega^2$ are real-valued, and $\omega$ is purely imaginary. In this case, the outer problem can be solved by a standard shooting procedure.
Namely, for a given test value of $\omega^2<0$, we numerically integrate Eq.~(\ref{eq:deltaphioutmode}) from $r=R$ to some $r=r_\textrm{out}$, with initial conditions on $r=R$ determined by the solution to the inner problem. Generically, the solution thus obtained diverges exponentially as $r \to \infty$. We then repeat this procedure by varying the value of $\omega^2$: unstable modes exist if a value of $\omega^2<0$ can be found so that $\delta \phi (r_\textrm{out}) \approx 0$ for sufficiently large $r_\textrm{out}$ (we typically use $\sim 100 R$).

For a stable mode, $\textrm{Im}(\omega) > 0$, and $Z^-(r)$ grows exponentially as $r \to \infty$. The frequencies $\omega$ are now complex since the operator is no longer Hermitian. To determine a solution via direct integration, we numerically solve Eq.~(\ref{eq:Z}) from $r=R$ to $r=r_\textrm{out}$ with initial conditions coming from the solution to the inner problem. We adjust the value of $\omega$ so that the solution thus obtained matches continuously to a solution of the form $Z^-(r)$ at $r_\textrm{out}$ [see Eq.~(\ref{eq:Zpm})]. For this purpose, we typically use $N=15$ in Eq.~(\ref{eq:Z}) and integrate up to $r_\textrm{out}\sim 30R-50R$. As we consider larger values of $r_\textrm{out}$, it is increasingly difficult to avoid contamination from the unwanted solution $Z^+(r)$, which decreases exponentially as $r \to \infty$, and the method becomes unreliable. The sensitivity of direct integration methods to numerical errors leads us to consider alternative procedures to solve the outer problem.

A method which turns out to be quite reliable involves the integration of Eq.~(\ref{eq:deltaphioutmode}) in the complex $r$-plane. This procedure was used in Ref.~\cite{Andersson1995b} in order to compute $w$-mode frequencies of stellar models in GR (see also \cite{Andersson1993} for the case of black hole modes). The basic idea is to avoid dealing with exponentially growing/decaying solutions by considering the problem in a region of the complex plane where the analytic continuation of the functions $Z^\pm$ have comparable amplitudes. Since $Z^\pm \propto e^{\pm i \omega r} =\exp \{\pm i [{\rm Re}(\omega){\rm Re}(r) - {\rm Im}(\omega){\rm Im}(r)]\} \exp \{\mp [{\rm Im}(\omega){\rm Re}(r) + {\rm Re}(\omega){\rm Im}(r)]\} $, as $|r| \to \infty$, the exponential divergence can be suppressed along a path in the complex plane that for large $|r|$ is a straight line with slope $-\textrm{Im}(\omega)/\textrm{Re}(\omega)$. We therefore set $r = R + \rho e^{i\theta}$, where $\rho$ is a real integration parameter (the distance to the stellar surface) and the constant phase angle $\theta$ is determined by
\begin{equation}\label{eq:theta}
\frac{\rho_\text{max} \sin \theta}{R + \rho_\text{max} \cos \theta} = - \frac{{\rm Im} (\omega)}{{\rm Re} (\omega)},
\end{equation}
where $\rho_\text{max}$ is taken to be $\sim 100 R$.
In most cases we can safely neglect $R$ in the denominator of the left-hand side of Eq.~(\ref{eq:theta}) and take $\tan \theta = -\textrm{Im}(\omega)/\textrm{Re}(\omega)$, but for modes with small real part we take instead $\cos\theta = -R/\rho_\text{max}$.

Concretely, we write Eq.~(\ref{eq:deltaphioutmode}) as 
\begin{eqnarray} 
&&\frac{d^2 \delta \phi}{d\rho^2} + e^{i\theta}\frac{1+e^{2\lambda_{(0)}}}{R+\rho e^{i\theta}} \frac{d\delta \phi}{d\rho} \nonumber \\
&&+e^{2i\theta} e^{2\lambda_{(0)}} \left(2 \Psi_{(0)}^2 + e^{- 2\nu_{(0)}} \omega^2\right) \delta \phi = 0,
\label{eq:complexplane}
\end{eqnarray}
where all background quantities are computed at $r = R + \rho e^{i\theta}$, and numerically integrate this equation from $\rho = \rho_\text{max}$ to $\rho = 0$. The boundary conditions at $\rho = \rho_\text{max}$ are determined from Eq.~(\ref{eq:Zpm}). We iterate in the quasinormal mode frequency $\omega$ until the exterior and interior solutions match at $\rho=0$ according to
\begin{equation*}
e^{-i\theta} \frac{1}{\delta \phi^{\rm out} (0)} \left. \frac{d\delta \phi^{\rm out}}{d \rho}\right|_{\rho =0} = \frac{1}{\delta \phi^{{\rm inn}} (R)} \left. \frac{d\delta \phi^{\rm inn}}{d r}\right|_{r=R}.
\end{equation*}
The integration of Eq.~(\ref{eq:complexplane}) benefits from the fact that in the vacuum region outside the star the exact form of the background quantities is known~\cite{Coquereaux}, and can be explicitly evaluated for complex values of $r$. However, the method outlined above can be used even when background quantities need to be computed numerically (sufficing to compute them along the desired path), which is the generic case in modified theories of gravity.

To compute QNMs of $\phi_{(0)}=0$ solutions, we also implemented a version of Leaver's continued fraction method \cite{Leaver1985}, following similar steps as in Ref.~\cite{Leins1993}. All methods give consistent results.

\subsection{Time domain approach}\label{sec:td}
In order to count on a consistency check of results obtained with the frequency domain techniques described above, here we follow an independent approach to extract oscillation modes of the second-order system~(\ref{eq:xi})-(\ref{eq:deltaphi}). The strategy is to pose an initial value problem for the perturbation fields $(\xi,\delta\phi)$ with appropriate boundary conditions, and employ numerical methods to determine their time evolution. Then, a spectral decomposition determines the vibration modes of the solution.
In practice, we reduce Eqs.~(\ref{eq:xi})-(\ref{eq:deltaphi}) to a first-order system, which we integrate in a finite, spherically symmetric spacetime domain $[0,t_\text{max}]~\times~[0,r_\text{max}]$, such that $r_\text{max}>R$. Initial data (at $t=0$) for the Cauchy problem are chosen of the form
\begin{eqnarray}
\delta\phi(0,r) &\propto& \exp\left[-(r - \bar{r})^2/\sigma^2\right], \label{eq:initial_deltaphi}\\
\delta\dot{\phi}(0,r) &=& 0,\label{eq:initial_dot_deltaphi}\\
\xi(0,r) &=& 0,\label{eq:initial_xi}\\
\dot{\xi}(0,r) &=& 0,\label{eq:initial_dot_xi}
\end{eqnarray}
for $r\in[0,r_\text{max}]$, with $\bar{r} \in (R,r_\text{max})$ and $\sigma \in (0,R)$. We typically use $(\bar{r},\sigma) \sim (9R, R/12)$. The gaussian profile in Eq.~(\ref{eq:initial_deltaphi}) corresponds to an initial scalar field perturbation consisting of a ``shell" placed in the exterior of the star, whereas the vanishing Lagrangian displacement in Eq.~(\ref{eq:initial_xi}) describes the initial absence of matter perturbations. Equations~(\ref{eq:initial_dot_deltaphi}) and~(\ref{eq:initial_dot_xi}) set the perturbations at rest in the initial time slice. In each subsequent time slice, we impose the solution to be regular at the origin and $\delta \phi (t,r)$ to satisfy an outgoing boundary condition.

Time evolution of the initial data is determined using the Method of Lines (see for instance Ref.~\cite{LeVeque-Book_FD}) with a third-order accurate Runge-Kutta integrator. Spatial derivatives are approximated by second-order accurate finite differences. The coordinates $(t,r)$ are discretized in a uniform grid of $(N_t+1, N_r+1)$ nodes $(t^n, r_i) = (n\Delta t, i\Delta r)$, with $(\Delta t, \Delta r) := (t_\text{max}/N_t, r_\text{max}/N_r)$, and $n \in \{ 0,1,2,...,N_t \}$, $i \in \{ 0,1,2,...,N_r \}$. In our simulations, usually $(N_t, N_r) = (2\times 10^6, 3.6\times10^4)$ and $r_\text{max} \sim 75R$. $\Delta t$ is determined by the Courant-Friedrichs-Lewy condition $\Delta t/\Delta r = \kappa_\text{CFL} < 1$, which we ensure by setting $\kappa_\text{CFL} \equiv 1/4$. We have verified that numerical solutions thus computed are self-convergent in the $L_2$-norm of relative errors, at least to second order.

\begin{figure}
\centering
\vspace{0.5cm}
\includegraphics[width=0.95\linewidth]{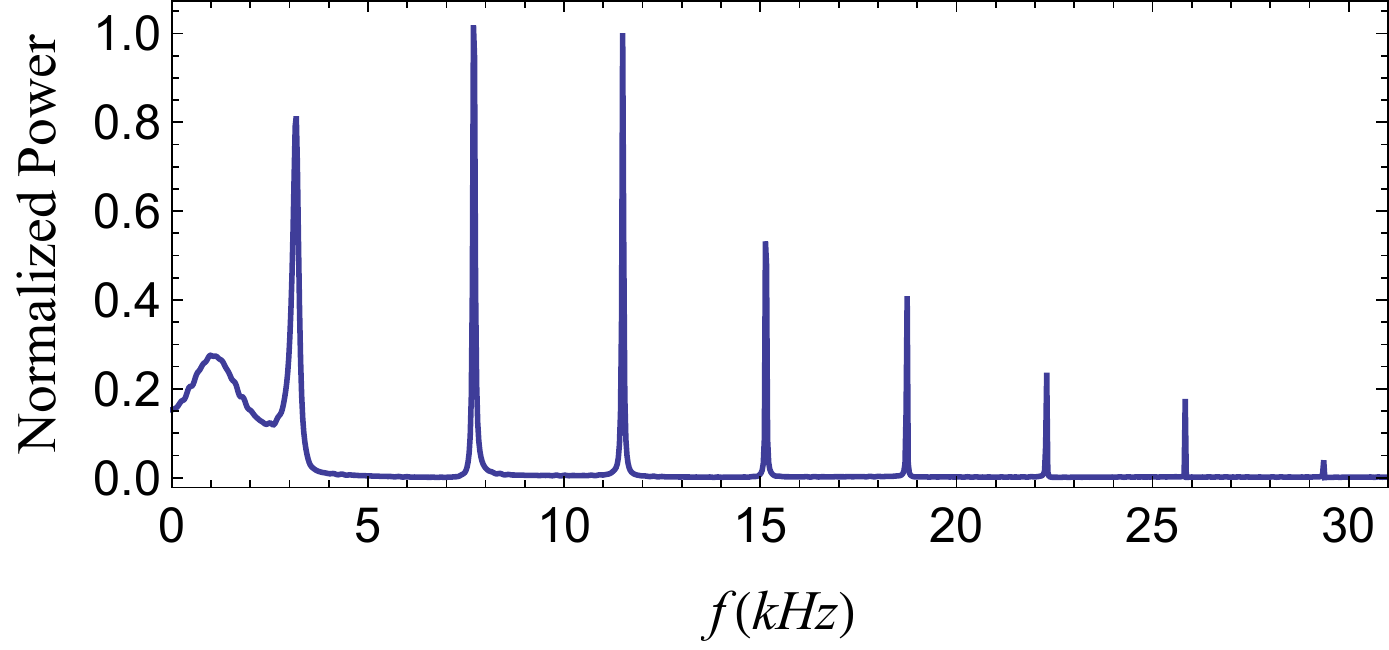}\;\;\;\;
\includegraphics[width=1.0\linewidth]{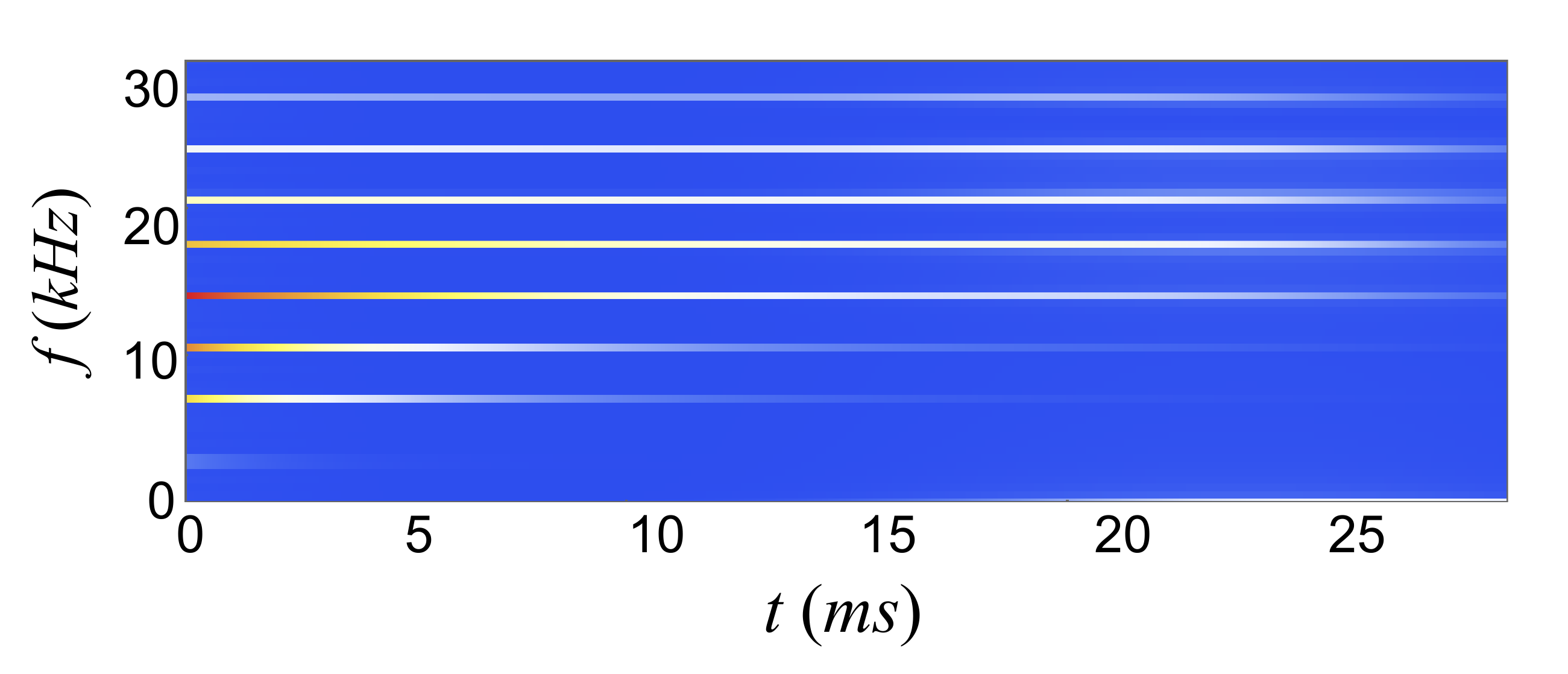}
\caption{Normalized power (upper panel) and spectrogram (lower panel) of scalar field perturbations sampled by a static observer outside a scalarized star with $\tilde{\rho}_c = 7.5\rho_0$ ($M/R \approx 0.237$) in M1 with $\beta = -5$. }
\label{fig:fourier}
\end{figure}

Schematically, as time evolution proceeds, the incoming part of the initial perturbation eventually reaches the background star, passes through it, and then propagates away. As the initial perturbation passes through the star, it generically excites its natural oscillation modes. We extract the corresponding oscillation frequencies by performing a spectral decomposition of a discrete time sampling of scalar field perturbations $\delta\phi_\text{obs}(t) := \delta\phi(t,r_\text{obs})$ at a fixed position $r_\text{obs} = 15R$. We sample $\delta\phi_\text{obs}(t)$ at a constant rate $1/f_s$. In our simulations, $(t_\text{max}, f_s) \sim (40\text{ms}, 10^6\text{Hz})$. Spectral decomposition of the time series data $\delta\phi_\text{obs}(t)$ is based on a Fourier analysis. Damping factors are extracted from the fitting parameters of Lorentzian functions in the vicinity of the peaks of Discrete Fourier Transforms (DFTs). For the sake of joint visualization of the relative power and damping of excited modes, we generate spectrograms of the data $\delta\phi_\text{obs}(t)$ using DFTs with a Blackman window function, maximizing resolution in frequency. Figure~\ref{fig:fourier} displays a representative example, showing that the full frequency spectrum consists of several overtones.

The frequency uncertainty intrinsic to our spectral analysis is $1/t_\text{max}$. However, numerical errors in the time domain integration can produce larger uncertainties. Therefore, error bars in Figs.~\ref{fig:QNM_beta-5} and~\ref{fig:QNM_beta100} for a given oscillation mode are estimated by taking the absolute difference of frequencies and damping times extracted from time domain simulations with two consecutive numerical resolutions $\Delta r$, $\Delta r/2$. 


\end{document}